\documentclass[%
reprint,
 amsmath,amssymb,
 aps,
]{revtex4-2}

\usepackage{graphicx}
\usepackage{dcolumn}
\usepackage{bm}

\usepackage{xcolor}

\begin{document}

\preprint{APS/123-QED}

\title{Elastic wave manipulation in geodesic lenses}%

\author{Matteo Mazzotti}
\author{Mohit Gupta}%

\author{Massimo Ruzzene}%
\affiliation{%
 P. M. Rady Department of mechanical Engineering, University of Colorado Boulder
}%

\author{Christian Santangelo}%
\affiliation{%
 Department of Physics, Syracuse University
}%

\date{\today}

\begin{abstract}
%
We investigate wave propagation in curved, thin elastic waveguides, where curvature is shown to be equivalent to a spatially modulated refractive index. We establish this relationship within a theoretical framework that leverages non-Euclidean transformations. Theoretical predictions illustrate how curved profiles can be designed to manipulate wave trajectories in analogy with geometrical optics principles. Experiments on 3D printed curved membranes confirm the theoretical predictions, and suggest curvature as a viable approach for guiding, focusing and steering of elastic waves.  
\end{abstract}

\maketitle



Since the pioneering works of Pendry \cite{Pendry2006} and Leonhardt \cite{Leonhardt2006}, transformation optics (TO) has been successfully employed for wave manipulation in various fields including electromagnetism, acoustics and mechanics \cite{chen2010,sun2017,McCall2018}. Application of TO concepts has enabled exotic effects such as cloaking \cite{schurig2006,Farhat2008a,Zhang2012a,Colombi2016}, superlensing \cite{Tsang2008,Wang2017a} and negative refraction~\cite{chen2010}. At the basis of TO is a conformal coordinate transformation that warps the waveguide geometry through material transformations that modulate the refractive index in space, leading to strongly anisotropic distributions. These refractive index distributions are commonly implemented by employing metamaterial concepts~\cite{padilla2006,schurig2006,valentine2008,alitalo2009}, which often impose manufacturing constraints and imply the ability to only realize discrete approximations of continuous distributions. These limitations, along with the intrinsic dynamics these concepts leverage, bound the range of achievable refractive indexes and the effective frequency ranges and bandwidths of operation.
 
An alternative approach to classical TO is offered by non-Euclidean transformations that map a planar surface into a curved space~\cite{McCall2018,MitchellThomas2018}. Using non-Euclidean transformations, a planar in-homogeneous waveguide can be mapped onto a curved homogeneous one where the Gaussian curvature plays the role of the refractive index~\cite{Leonhardt,Kamien2009,schultheiss2010,Bekenstein2017,Delphenich2020}. This allows mimicking spatially varying material properties in a planar waveguide purely via geometric transformations that induce curvature. The approach was first proposed by Rinehart in constructing the geodesic equivalent of a planar Luneburg lens~\cite{rinehart1948,Rinehart1952}. Since then, non-Euclidean transformations have been proposed in theoretical and experimental studies of the Luneburg's problem,~ \cite{Southwell77,Sochacki86,Sochacki1988,Sarbort2012,MitchellThomas2014,MitchellThomas2018} the analysis of geodesic lenses with an equivalent infinite refractive index such as Eaton and Maxwell fish-eye lenses,~\cite{Sarbort2012,Horsley2014} the implementation of conformal optical devices for perfect imaging,~\cite{Minano_2010,Xu2019a,Xu2019b} and the design of multibeam antennae.~\cite{Liao2018,Fonseca2020}

In spite of the considerable developments in optics and electromagnetics, elastic geodesic lenses have received little attention to date. Prior studies in this area include the work of Evans et al.~\cite{Evans2013}, which has investigated flexural waves in thin anisotropic curved shells of complex geometry, and has leveraged their bi-refringent behavior to achieve total wave reflection at interfaces between regions of negative and positive Gaussian curvature. More recently, the 2D elastic analog of a wormhole was presented in~\cite{Zhu2018a} through numerical and experimental studies that demonstrate that the required infinite refractive index along the axis of the wormhole, unattainable via material anisotropy, can instead be achieved through geometric curvature alone. 
This equivalence enables the study of the wave dynamics within the wormhole on a virtual Euclidean plane by employing classical ray tracing methodologies. In addition, the elastic analog of the event horizon is also studied by means of geodesic analysis, which establishes the minimum radius from the center of the wormhole required for a flexural ray to avoid trapping.
Other related work involving the modification of the geometric profile of a homogeneous elastic waveguide investigated the implementation of acoustic black holes~\cite{Krylov2014,Zhao2014a,Zhu2015a,Pelat2020}, cloaks~\cite{Farhat2009,Stenger2012,Zhao2020a} and singular lenses~\cite{Lee2021}. All these studies consider rotationally symmetric structures, for which the equivalence between planar refractive index and geometric curvature is now well established~\cite{MitchellThomas2014}.

The objective of this work is to theoretically and experimentally investigate the geometric equivalence between 2D elastic gradient index (GRIN) waveguides and their geodesic counterparts for general distributions of refractive index and Gaussian curvature. The development of a non-Euclidean transformation framework for general geometries offers numerous opportunities for the design of curved surfaces capable of wave focusing, guiding and steering which enable functionalities that go beyond those achieved by rotationally symmetric structures. The equivalent refractive index properties obtained through curvature can be induced without the need for potentially complex patterning of material and/or geometric features, as for example in the case of metamaterials-based GRIN concepts~\cite{Jin2016,Jin2019,Zhao2016,Danawe2020,Chen2016,Tol2016,Tol2017}.

In this letter, we first present the theoretical background for the analysis of flexural waves in elastic shell-like waveguides of generic curvature. We describe their behavior in the short wavelength limit (\emph{i.e.} when the wavelength is small compared to the characteristic radii of curvature), and we introduce  a non-Euclidean transformation that enables wave analysis in terms of geodesics and ray tracing. We then illustrate these concepts through numerical simulations and experiments conducted on 3D-printed waveguides with selected curvature profiles. The results highlight that the experimentally measured focal regions and caustic networks are predicted with good accuracy by the profile geodesics as ray trajectories.


\begin{figure}
\includegraphics[width=0.48\textwidth]{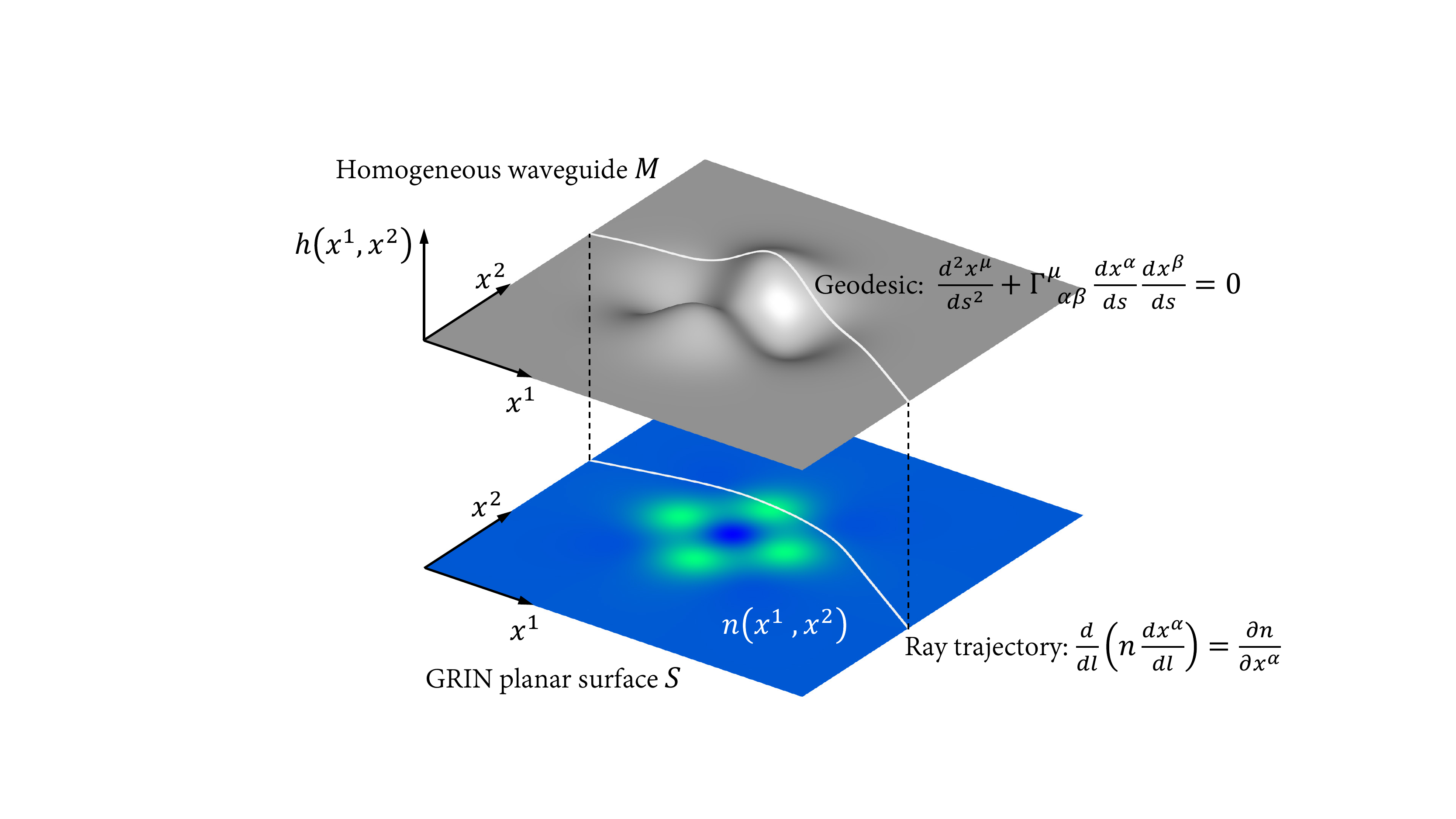}
\caption{Mathematical model of the GRIN planar surface $S$ and the geometrically equivalent homogeneous waveguide $M$.}
\label{fig:model}
\end{figure}

We begin by deriving the relation between the trajectories of elastic waves propagating on flat GRIN surfaces and on curved surfaces in the geometrical optics limit. 
To this end, we first consider the planar waveguide $S$ of variable refractive index $n(x^{1},x^{2})$ shown in Fig.~\ref{fig:model}, whose mid-surface lies on the Euclidean plane $x^{1},x^{2}$. The wave trajectories on $S$ are described by the classic ray equation~\cite{sharma82}
\begin{equation}
\frac{d}{dl}\left( n \frac{dx^{\alpha}}{dl}\right) = \frac{ \partial n}{\partial x^{\alpha}}, \quad \alpha=1,2,
\label{eq:ray1}
\end{equation}
where $dl^2=\delta_{\alpha\beta} dx^{\alpha} dx^{\beta}$ is the arc length in Euclidean metric and $\delta_{\alpha\beta}$ denotes the Kronecker's delta. On $S$, the infinitesimal path length is given by $da^2 = n^2(x^{1},x^{2}) dl^2$.

Next, we look for a curved waveguide $M$ of homogeneous material, constant thickness $t$, and midsurface profile $h(x^{1},x^{2})$ (Fig.~\ref{fig:model}) which supports the propagation of elastic waves along trajectories that coincide with those given by Eq.~\eqref{eq:ray1}. We will denote this waveguide as geometrically equivalent to $S$. 
Without loss of generality, we consider $M$ as an arbitrarily curved thin elastic shell undergoing pure bending deformation. The governing equation of motion for this geometry can be found in~\cite{norris,norris2}, within the assumption $|k| \gg (rt)^{-1/2}$, where $r(x^{1},x^{2})$ is the characteristic radius of curvature, while $k$ is the flexural wavenumber.  This condition is typically verified when wavelengths $\lambda=k^{-1}$ are much shorter than the characteristic scale of the curved shell and much longer than the thickness $t$~\cite{Germogenova1973}. In this wavelength regime, the dispersion relations for flexural waves is given by~\cite{norris,norris2}
\begin{equation}
	\frac{\omega^2}{c_p^2} - \frac{t^2}{12} \left( k_\alpha k^{\alpha} \right)^2 - (1-\nu^2) \left( \frac{k_{\alpha} k_\beta \epsilon^{\alpha \gamma} \epsilon^{\beta \rho} b_{\gamma \rho}}{k_\alpha k^\alpha} \right)^2 = 0,
\label{eq:dispersion}	
\end{equation}
where $\omega$ is the circular frequency, $\nu$ is the Poisson ratio, $c_{p}=\omega/k$ is the phase velocity of the flexural wave, $\epsilon^{\alpha \beta}$ is the antisymmetric Levi-Civita tensor,~\cite{Carmo} and $k_\alpha(s)$ are treated as the canonical momenta of the ray coordinates. The flexural rays, described by trajectories $x^\alpha(s)$, can be derived by considering the Hamiltonian for the shell, which is given by~\cite{norris}
\begin{equation}
H = \frac{1}{2 \omega} \frac{c_p^2 t^2}{12} \left[ k_\alpha(s) g^{\alpha \beta}(x(s)) k_\beta(s) \right]^2, 
\label{eq:H}
\end{equation}
where $g^{\alpha\beta}$ is the contravariant metric tensor~\cite{Carmo}. In the case of a Monge patch, its covariant counterpart $g_{\alpha\beta}$ describes the metric $ds^2 = g_{\alpha\beta} dx^{\alpha} dx^{\beta}$ of $M$, and is expressed as
\begin{equation}
g_{\alpha\beta}=\delta_{\alpha\beta}+\partial_{\alpha}h \partial_{\beta}h. 
\label{eq:gabm}
\end{equation}
The wavenumber $k_\alpha(s)$ in Eq.~\eqref{eq:dispersion} can be obtained by employing the first fundamental relation of Hamiltonian dynamics $\dot{x}^\alpha(s) = \partial H / \partial k_\alpha$, which gives
\begin{equation}
	k_\alpha = \left( \frac{3 \omega}{c_p^2 t^2} \right)^{1/3} \frac{ g_{\alpha \beta} \dot{x}^\beta}{(\dot{x}^2)^{1/3}}.
	\label{eq:kalpha}
\end{equation}
Substituting Eq.~\eqref{eq:kalpha} into the second fundamental relation $\dot{k}_\alpha(s) = -\partial H / \partial x_\alpha$, and letting $v^\beta \equiv \dot{x}^\beta/(\dot{x}^2)^{1/3}$, yields
\begin{equation}
2 \frac{1}{(\dot{x}^2)^{1/3}} g_{\gamma \beta} \dot{v}^\beta + 2 \partial_\alpha g_{\gamma \beta} v^{\alpha} v^\beta -  \partial_\gamma g_{\alpha \beta} v^\alpha = 0,
\label{eq:geo0}
\end{equation}
where all temporal dependence has been eliminated. Replacing $x^\alpha(s)$ back into Eq.~\eqref{eq:geo0} gives
\begin{equation}
\label{eq:geodesic}
\frac{1}{(\dot{x}^2)^{1/3}}  \frac{d}{ds} \left( \frac{\dot{x}^\mu}{(\dot{x}^2)^{1/3}} \right) + \Gamma^\mu_{~ \alpha \beta} \left( \frac{\dot{x}^\alpha}{(\dot{x}^2)^{1/3}} \right) \left( \frac{\dot{x}^\beta}{(\dot{x}^2)^{1/3}} \right) = 0,
\end{equation}
where 
\begin{equation}
\Gamma^\mu_{~\alpha \beta} = \frac{1}{2}g^{\mu \nu} \left( \partial_\alpha g_{\beta \nu} + \partial_\beta g_{\alpha \nu} - \partial_\nu g_{\alpha \beta} \right)
\label{eq:christoffel1}
\end{equation}
are the Christoffel symbols of the second kind. 

The parameter $s$ is arbitrary. To simplify Eq.~\eqref{eq:geodesic}, we reparameterize the rays in terms of their arc length, $l$, defined by $\dot{x}_\alpha g^{\alpha \beta} \dot{x}_\beta = 1$, for which Eq.~\eqref{eq:geodesic} reduces to the classic geodesic equation,
\begin{equation}
\frac{d^2 x^{\mu}}{ds^2} + \Gamma^\mu_{~\alpha \beta} \frac{d x^{\alpha}}{ds} \frac{d x^{\beta}}{ds} = 0,
\label{eq:geo1}
\end{equation}
Geodesics are loosely defined as shortest paths on $h(x^{1},x^{2})$ followed by a particle traveling at constant tangential speed~\cite{Leonhardt}. Therefore, within geometrical optics limits and owing to Fermat's principle, the trajectory on $M$ of flexural waves propagating at phase speed $c_{p}$ satisfies Eq.~\eqref{eq:geo1}. 

We now map the GRIN planar waveguide $S$ onto a geometrically equivalent curved waveguide $M$, specifically in order to establish the relation between the refractive index $n(x^{1},x^{2})$ and the geometric profile  $h(x^{1},x^{2})$. 
This relationship implies that geodesic curves on $M$ are mapped onto ray trajectories on $S$.
This is achieved by imposing that $da=ds$, i.e.
\begin{equation}
    g_{\alpha\beta} dx^{\alpha} dx^{\beta} = n^2 \delta_{\alpha\beta} dx^{\alpha} dx^{\beta}.
    \label{eq:length}
\end{equation}

Equation~\eqref{eq:length} can be directly applied when the metric tensor is given in local isothermal coordinates $(u,v)$ such that $g_{u v}=\phi^2\delta_{u v}$ \cite{Leonhardt}, allowing us to identify the conformal factor $\phi^2$ with $n^2$.
However, while $(x^{1},x^{2})$ are isothermal coordinates for the refractive index, they usually are not for the Monge patch. 
Consequently, finding the conformal factor of the Monge patch leads to a refractive index described with respect to a set of planar isothermal coordinates $(u,v)$ that do not correspond to $(x^{1},x^{2})$, and a direct mapping between $h(x^{1},x^{2})$ and $n(x^{1},x^{2})$ would require an additional coordinate transformation for the refractive index from $(u,v)$ back to $(x^{1},x^{2})$. The approach based on the use of the conformal factor can be avoided by resorting to the Gaussian curvature $K$, which is invariant with respect to a parameterization of a surface. Specifically, we can use this property to enforce the waveguides $S$ and $M$ to be geometrically equivalent by imposing their Gaussian curvature at a generic coordinate $(x^{1},x^{2})$ to be identical.

%
Using the definition of $K(u,v)$ in isothermal parameterization
\begin{equation}
    K(u,v) = - \frac{1}{2 \phi^2(u,v)} \left( \frac{\partial^2}{\partial u^2} + \frac{\partial^2}{\partial v^2} \right) \ln \phi^2(u,v), 
\end{equation}
to express the Gaussian curvature as a function of the refractive index $n(x^{1},x^{2})$, and equating it to the expression of $K(x^{1},x^{2})$ in terms of the surface elevation $h(x^{1},x^{2})$ \cite{Carmo},
we finally obtain the mapping
\begin{equation}
    - \frac{\nabla^2 \ln n^2(x^1,x^2)}{2 n^2(x^1,x^2)} 
    = \frac{\mathrm{det}~[\partial^2 h(x^1,x^2)/\partial x^{\alpha} \partial x^{\beta}]}{[1+(\nabla h(x^{1},x^{2}))^2]^2}
    \label{eq:ntoh}
\end{equation}

Equation~\eqref{eq:ntoh} defines the local mapping between the refractive index $n(x^{1},x^{2})$ and $h(x^{1},x^{2})$ through the Gaussian curvature $K(x^{1},x^{2})$, and can be regarded as the two-dimensional generalization of the non-Euclidean transformations described in \cite{Doric1983,Sarbort2012,MitchellThomas2014,Horsley2014,MitchellThomas2018} for rotationally symmetric curved surfaces. 

\begin{figure}
\includegraphics[width=0.48\textwidth]{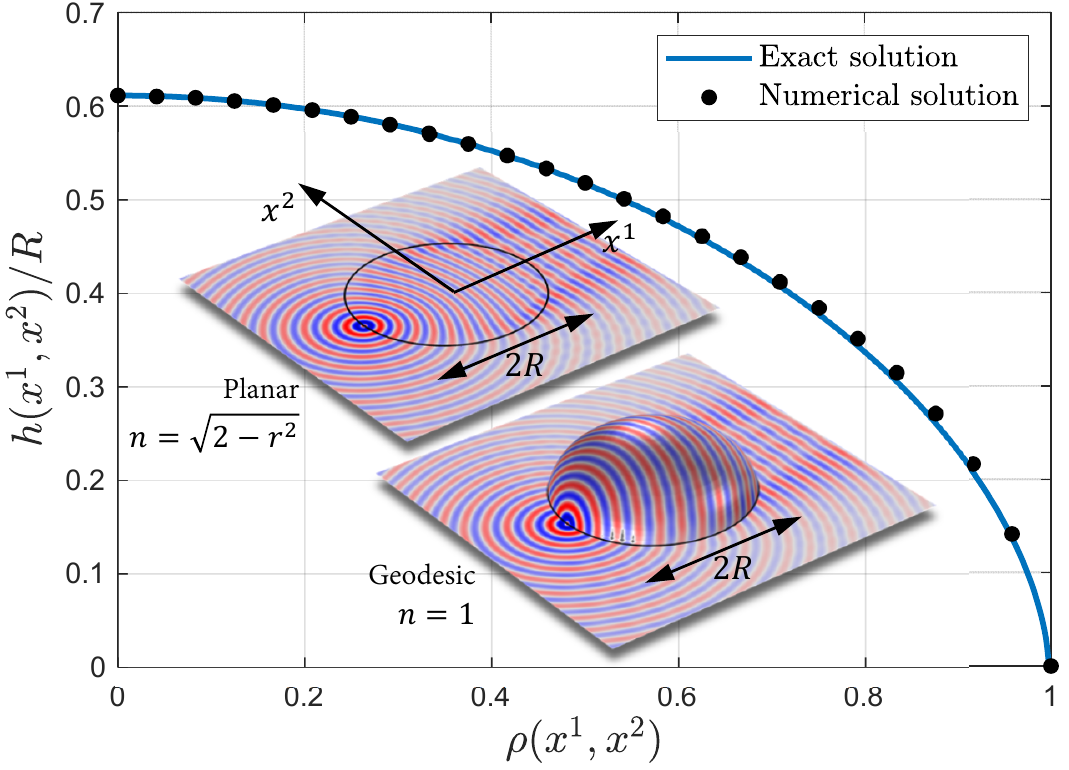}
\caption{Elevation profile of a geodesic Luneburg lens with unit radius and comparison of the wavefield generated by a monopole source with respect to the flat case.}
\label{fig:luneburg}
\end{figure}

To illustrate this mapping, we show how Eq.~\eqref{eq:ntoh} can be applied to the case of a planar Luneburg lens~\cite{rinehart1948,Rinehart1952} of radius $R=1$ and refractive index $n=(2-\rho^2)^{1/2}$, where $\rho=(x^{1}+x^{2})^{1/2} R^{-1} \in [0,1]$. This lens can be transformed into the equivalent geodesic version $M$ by solving Eq.~\eqref{eq:ntoh} in terms of $h(x^{1},x^{2})$ over a circular domain of radius $R=1$, with Dirichlet boundary conditions $h(R=1)=0$. The resulting profile $h(\rho)$ is shown in Fig.
Fig.~\ref{fig:luneburg}, where analytical predictions obtained in \cite{MitchellThomas2014,Liao2018} (solid line) are compared to those predicted through the numerical discretization of Eq.~\eqref{eq:ntoh} (filled circles), the details of which are provided in the Supplementary Material.
Figure~\ref{fig:luneburg} also compares the harmonic wavefields of wavelength $\lambda=R/5$ generated by a monopole source located on the boundary of the two lenses: the close similarity demonstrates the geometric equivalence between surface shape and the refractive index. 

\begin{figure*}
\includegraphics[width=0.99\textwidth]{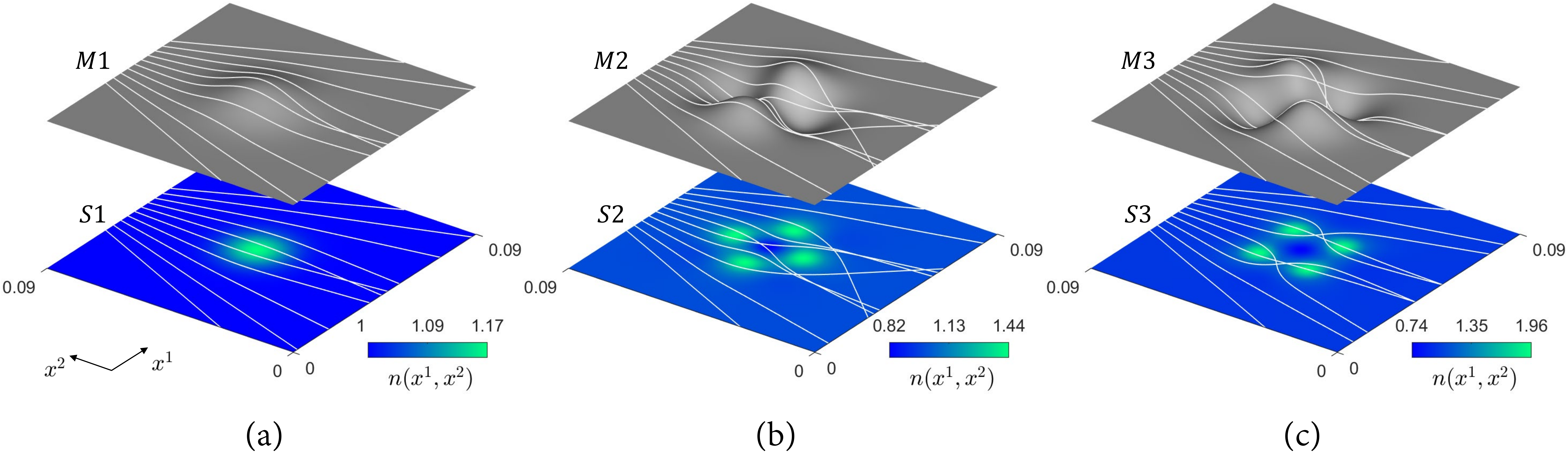}
\caption{\label{fig:waveguides}Modulated Gaussian profiles with geodesics and corresponding GRIN planar surfaces with ray trajectories. The profiles are generally expressed as $h(x^{1},x^{2})=m(x^{1},x^{2}) A \textrm{exp}[(((x^{1})^{2}+(x^{2}))^{2}/R^{2})/2]$, with: (a) $M1$: $A=6.0$ mm, $R=5.08$ mm, $m=1$; (b) $M2$ $A=850$ mm, $R=8.89$ mm, $m=[(x^{1})^{2}-(x^{2})^{2}]/R$; (c) $M3$ $A=2000$ mm, $R=8.89$ mm, $m=x^{1}x^{2}/R$.}
\end{figure*}
\begin{figure*}
\includegraphics[width=0.99\textwidth]{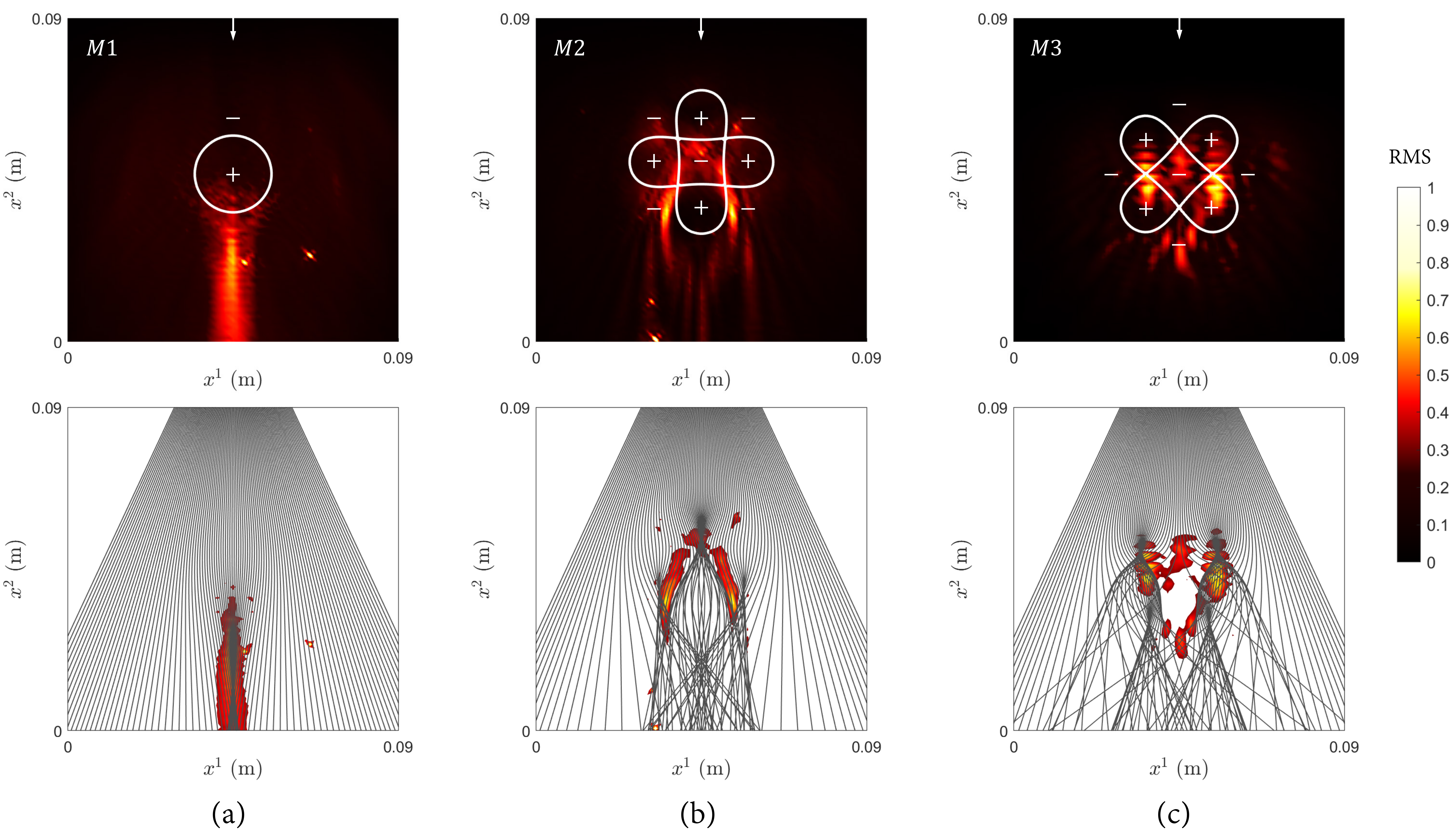}
\caption{\label{fig:results}Root Mean Square (RMS) maps of the out-of-plane velocity component of the experimental $A_{0}$ Lamb mode with contour lines of zero Gaussian curvature (top view) and geodesics (bottom view) of the membranes (a) $M1$, (b) $M2$ and (c) $M3$. All the RMS maps are normalized with respect to their maximum values. The white arrow in the RMS maps indicates the direction of the incident $A_{0}$ Lamb mode.}
\end{figure*}

To demonstrate the general validity of Eq.~\eqref{eq:ntoh}, we analyze the waveguiding properties of three different surfaces. In particular, we demonstrate that the locations at which guided waves focus match the ray envelopes (caustics) predicted through geodesic analysis, and that such distribution can also be obtained by ray tracing on a geometrically equivalent GRIN planar surface determined from Eq.~\eqref{eq:ntoh}.
To this end, we consider the three curved surfaces $M1$, $M2$ and $M3$ in Fig.~\ref{fig:waveguides}. 
Each curved surface was transformed into its geometrically equivalent GRIN planar counterpart by performing the numerical solution of Eq.~\eqref{eq:ntoh} as indicated in the Supplementary Material. 
The refractive index maps obtained from the profiles $M1$, $M2$ and $M3$ are shown in Figs.~\ref{fig:waveguides}(a-c).
Geodesics for different incident directions were computed on the curved surfaces using the method outlined in~\cite{burazin}, while the corresponding ray trajectories were traced on the GRIN planar surfaces by using the algorithm described in~\cite{sharma82}.
These results confirm that the same waveguiding effect is obtained in the curved and planar two-dimensional spaces by either modulating the geometric profile $h(x^{1},x^{2})$ or the refractive index $n(x^{1},x^{2})$ according to Eq.~\eqref{eq:ntoh}.

The procedure and the ability of geodesic and ray analyses to predict wave propagation and the existence of focal points and caustic networks was tested numerically and experimentally on 3D printed membranes of $1.27$ mm thickness reproducing the profiles $M1$, $M2$ and $M3$. In the experiments, Lamb waves were generated on the membranes by a single circular piezoelectric actuator excited by a chirped signal in the $35-300$ kHz range. The out-of-plane component of the resulting guided wavefiled was recorded using a Laser Doppler vibrometer and filtered in the wavenumber domain to retain only the fundamental antisymmetric $A_{0}$ mode in the wavelength range $\lambda \in [2.75,5.70]$ mm (see Supplementary Material for details on the experimental set-up).

The results obtained from the experimental tests are shown in Fig.~\ref{fig:results}, where the RMS of the out-of-plane velocity component of the experimental $A_{0}$ Lamb mode (top view) is compared to a dense distribution of geodesics (bottom view). In the RMS maps, the arrow indicates the direction of propagation of the incident $A_{0}$ mode, while the continuous lines represent the loci of zero Gaussian curvature.  
For the membrane $M1$, the shape of the Gaussian profile induces a distortion of the guided wavefield that results in an elongated focal line (top view of Fig.~\ref{fig:results}(a)). 
This focal line corresponds to the caustic network formed by the geodesics on $M1$ (bottom view of Fig.~\ref{fig:results}(a)) and, in the view in Fig.~\ref{fig:waveguides}(a), to the ray trajectories on $S1$.
The experimental RMS of the membrane $M2$ (Fig.~\ref{fig:results}(b)) shows a more complex pattern, with two curved focal lines departing from the center of the Gaussian profile. Also in this case, it is possible to note that the locations at which flexural waves focus (red color) are intercepted by caustics formed by the geodesics. Consistently with Fermat's principle and the observations in~\cite{Arnott1989}, these caustics appear to be approximately aligned with the virtual boundaries that separate regions with $K>0$ (indicated with the ``$+$'' symbol) from regions with $K<0$ (``$-$'' symbol). Along these boundaries, the two-dimensional space is locally flat, which minimizes the time of flight between two locally close points.
A similar behavior is also observed for the waveguide $M3$ (Fig.~\ref{fig:results}(c)), in which the caustics form two distinct focal points at the loci where contours of zero curvature intersect, and that are in good agreement with the experimental observations. It is also interesting to note that, near the center of the waveguide, the ray trajectories exhibit a strongly diverging behavior as they pass through a region of negative Gaussian curvature. Based on the distribution of $n(x^{1},x^{2})$ obtained from Eq.~\eqref{eq:ntoh} (Fig.~\ref{fig:results}(c)), the equivalent refractive index on this region reaches a minimum of $n=0.74$ at the center of the waveguide and increases rapidly to a maximum of $n=1.96$ while moving towards the center of the neighboring modulated Gaussian domes, thus exhibiting a large spatial gradient. 
The diverging behavior of ray trajectories in this region offers a simple interpretation of the gradient of the refractive index $\partial n/\partial x^{\alpha}$ in Eq.~\eqref{eq:ray1}, which acts as an equivalent force that locally deflects a particle (viz., a ray trajectory) from a region of lower to a region of higher refractive index~\cite{Liu2019a}. 
By comparing the maps of $n(x^{1},x^{2})$ in Fig.~\ref{fig:waveguides} with those of $K(x^{1},x^{2})$ in Fig.~\ref{fig:results}, we can also draw the equivalent conclusion
that flexural rays propagating from regions of negative (positive) Gaussian curvature to regions of positive (negative) Gaussian curvature will tend to converge (diverge). In general, caustics will be expected to form near regions in which the Gaussian curvature vanishes~\cite{Kamien2009}.

In conclusion, the theoretical and experimental investigations performed in this study form the basis for the design of curved profiles that mimic a variable distributions of the refractive index in a two-dimensional flat space, and enable the realization of wave focusing and steering effects in analogy to geometrical optics principles. Our results suggest the Gaussian curvature as an attractive alternative to strategies based on the local tailoring of material properties and geometrical patterns that have gained in popularity for gradient-index lens design.
Being based on purely geometric arguments, the concepts presented in this work can also be extended to any problem of geometrical optics and electromagnetism, where the design of geodesic lenses from their planar GRINS counterparts has been mainly limited so far to rotationally symmetric geometries.


\providecommand{\noopsort}[1]{}\providecommand{\singleletter}[1]{#1}%

\end{document}